\documentstyle[aps,epsfig]{revtex}
\begin{document}
\draft
\title{Single-Particle Properties of a Two-Dimensional Fermi Liquid 
at finite Frequencies and Temperatures}

\author{Jungsoo Kim and D. Coffey}

\address{Department of Physics, State University of New York, 
Buffalo, NY 14260}

\date{\today}
\maketitle

\begin{abstract}
We review the leading momentum, frequency and temperature dependences
 of the single particle self-energy and the corresponding term in 
the entropy of a two dimensional Fermi liquid (FL) with a free 
particle spectrum. We calculate the corrections to these leading 
dependences for the paramagnon model and the electron gas and
 find that the leading dependences are limited to regions of energy 
and temperature which decrease with decreasing number density of 
fermions. This can make it difficult to identify the frequency and 
temperature dependent characteristics of a FL ground state in 
experimental quantities in low density systems even when 
complications of band structure and other degrees of freedom are 
absent. This is an important consideration when the normal state 
properties of the undoped cuprate superconductors are analyzed.
\end{abstract}

%
%

\section{Introduction}
 The 2D nature of the cuprate oxide superconductors and the fact
 that their normal state properties do not exhibit leading FL 
behavior have lead to the suggestion that the ground state is not a 
FL. In a FL the resistivity should follow a $T^2$ dependence as the 
temperature $T$ goes to zero in the absence of impurities and it has
 been argued that the absence of this characteristic 
FL behavior in several experiments for the cuprate 
superconductors\cite{Iye,Takagi,Ando,Timusk} is evidence 
that there is a qualitative difference between 
the normal state of cuprates and that in the other metallic system. 
Anderson\cite{Anderson} has argued that the ground 
state of cuprates is 
close to the one-dimensional system in which the elementary 
excitations are collective modes, spinons and holons, with spin and 
charge degrees of freedom decoupled as in the 1D Luttinger liquid. 
Varma \emph{et al}.\cite{Varma} have put forward a model in which 
the 
weight in the quasiparticle pole vanishes logarithmically at 
$p=p_f$. This is similar to case of the Luttinger liquid although in
 that case the weight is the quasiparticle poles at the Fermi 
surface vanishes as $z_F\sim (p-p_f)^\alpha, 1>\alpha>0$\cite{Voit}. 
The stability of the FL ground state has been investigated in 
perturbation theory\cite{Engelbrecht,Fukuyama,Serene,Coffey}, by 
renormalization group calculations\cite{Shanka} and bosonization of 
the fermions\cite{Houghton,Castro}. These approaches have not 
revealed any sign of FL breakdown other than due to the 
instabilities of FL ground state familiar from three-dimensional
(3D) case : BCS, CDW or SDW. Metzner \emph{et al.} have reviewed 
work on the role 
of strong forward scattering in determining the ground state of Fermi 
systems in different dimensions.\cite{Metzner}
As they point out for short interactions or the Coulomb interaction
the properties of Fermi systems in dimensions greater than one
are those of usual Fermi liquid.
 More recently Castellani \emph{et 
al}.\cite{Castellani} have 
suggested that the anomalous normal state properties of the cuprates
 are due to quantum critical points associated with 
antiferromagnetism and charge density wave instabilities in 
different doping ranges. In the region of a quantum critical point 
temperature provides the only energy scale which it is argued is 
consistent with data on the cuprates.

 An alternative explanation is that, although the ground state 
develops from FL at $T=0$, the characteristic FL temperature 
dependence is restricted to low temperatures compared to the high 
superconducting transition temperature, $T_c$, or compared to small 
energy scales given by large variations in the density of states 
due to bandstructure effects such as a Van Hove 
singularity\cite{Aristov}.

 The energy and temperature dependent characteristics of a FL arise 
from quasiparticle interactions. Corrections 
to these dependences also
 arise from the quasiparticle interactions themselves at energy and 
temperature scales which depend on the nature of interactions. The 
corrections limit the characteristic FL dependences to the vicinity 
of the Fermi surface and to low temperature even in the absence of 
bandstructure and other effects. We investigate the 
corrections to these leading FL energy and temperature dependences 
using a parabolic band in order to isolate 
the effects of bandstructure. It is useful to 
investigate the leading corrections for the FL with a simplest 
model where non-essential complications are absent. Here we 
calculate $\Sigma'(p,E)$ and $\Sigma''(p,E)$ for the paramagnon 
model\cite{Doniach} and the electron gas. The paramagnon model 
describes the physical system close to 
a ferromagnetic instability where the self-energy contribution comes
 from incoherent long-wavelength spin fluctuations through particle-
hole($ph$) channel. The closer the system is to the instability the 
stronger are the leading corrections to FL behavior 
since they come from the 
long wavelength limit of the effective interaction. This makes the 
paramagnon model ideal for investigating corrections and the regions 
of energies and temperatures over which they can characterize 
calculated quantities. 
Since the 2D electron liquid has recently been used to investigate 
the corrections to FL behavior, we also calculate these corrections 
for the Coulomb interaction. In these calculations $\Sigma(p,E)$ is 
approximated by the $ph$ channel contribution.
 Unlike 3D this channel remains 
important at low densities in 2D so that our results can also be 
applied to low densities. The difference between 2D and 3D  
is that the density of states for a parabolic band 
at the Fermi surface is independent of 
density in 2D. Consequently  repeated scattering in the $ph$ 
channel remains important at low density in 2D\cite{Kim}. 

 In the section II we introduce the paramagnon model and in the 
section III discuss the results for the single-particle self-energy 
and thermodynamic properties.  In the section IV we 
discuss the same kind of corrections for the 2D electron gas. We
give our conclusions in section V.

\section{Model}

 To calculate the leading corrections to the single-particle self-
energy  we first use a short range interaction between fermions. 
The Hamiltonian is 
\begin{equation}
H=\sum_{{\mathbf p}, \sigma} \xi_{{\mathbf p}} 
c^\dagger_{{\mathbf p},\sigma}c_{{\mathbf p},\sigma}+ 
\sum_{{\mathbf p},{\mathbf q},\sigma, \sigma\prime}I({\mathbf q})
 c^\dagger_{{\mathbf p},\sigma}c_{{\mathbf p} ',\sigma\prime}
c^\dagger_{{\mathbf p} ' -{\mathbf q},\sigma\prime}c_{{\mathbf p}
 +{\mathbf q},\sigma}
\end{equation}
where $\xi_p=(p^2 -p_f^2)/2m$, $m$ is the mass of the fermions, $p_f$
 is the Fermi momentum and $I$ is the 
strength of the interaction with a cutoff, $q_c$. This model, 
called the paramagnon model, was used by Engelsberg \emph{et al}.
\cite{Doniach} to calculate the corrections to the linear 
temperature dependences in specific heat of normal liquid $^3He$. 
 Within this model the leading corrections to FL behavior come from 
low energy long wavelength paramagnons. 
The importance of paramagnons 
in liquid $^3He$ 
was suggested by the enhancement of the observed static paramagnetic
 susceptibility and were shown to provide an explanation of the size 
of corrections to the linear temperature dependence of the specific 
heat. The single-particle self-energy is given by 
the repeated scattering of $ph$ pairs which leads to an effective 
interaction, $V^{eff}(\bf q, \omega)$.
\begin{equation}
\Sigma({\mathbf p}, \imath E_n)=-T\sum_{{\mathbf q}, 
\omega_l}G({\mathbf p} - {\mathbf q}, \imath E_n -\imath \omega_l)
{V}^{eff}({\mathbf q}, \omega_l)
\end{equation}
where $G({\mathbf p}, \imath E_n)$ is the unperturbed temperature 
Green function and $\omega_l$ is Bose Matsubara frequencies. 
$V^{eff}(\bf q,\omega)$ has two independent channels, the symmetric 
(s) and the 
antisymmetric (a) channels corresponding to spin exchanges of 0 or 1
 and is given by 
\begin{equation}
{V}^{eff}({{\mathbf q}},\omega)={1\over 2}{V^2_s\chi
({\mathbf q}, \omega)\over {1-V_s\chi ({\mathbf q}, \omega)}}+
{3\over 2}{V^2_a\chi
({\mathbf q}, \omega)\over {1-V_a\chi ({\mathbf q}, \omega)}}
\end{equation}
where $V_s=I$ and $V_a=-I$. $\chi({\mathbf q},\omega)$ is the 
polarization function for a 2D parabolic band\cite{Stern}.
 At $T=0$ the self-energy contributions 
from the real and imaginary part on shell ($E=\xi_p$) is 
\begin{eqnarray}
\Sigma({\mathbf p},\xi_{\mathbf p})&=& \sum_{\mathbf q}[\Theta(\xi_{
\mathbf p}-\xi_{{\mathbf p}-{\mathbf q}})-\Theta(-\xi_{{\mathbf p}-
{\mathbf q}})]V^{eff}({\mathbf q},\xi_{{\mathbf p}}-\xi_{{\mathbf p}
-{\mathbf q}})
\\ \nonumber \cr 
&+&\int^{\infty}_{-\infty}{d\eta \over {2\pi}}\sum_{\mathbf q}
G({\mathbf p}-{\mathbf q},\xi_{\bf p} -\imath \eta)V^{eff}
({\mathbf q},\imath \eta)
\end{eqnarray}
with $\chi({\mathbf q},\omega)=\chi'({\mathbf q},\omega)+\imath\chi''
({\mathbf q},\omega)$ in $V_{eff}({\mathbf q},\omega)$.
 The leading corrections in $\Sigma(p,E)$ of interest here comes from
the  long-wavelength limit and are contained in the first term since
 the second term, the line integral along the imaginary frequency 
axis, $\omega=\imath \eta$, does not contribute to the energy 
dependence in the long wavelength limit.  The line integral in the 
equation (4) does not contribute to the $\Sigma''(p,E)$ since the 
imaginary part of the integrand is odd in $\eta$ and 
gives contributions to $\Sigma'(p,\xi_p)$ which are proportional to 
$q_c^2$.

\section{Results}
\subsection{Self-Energy At Zero Temperature}

\begin{figure}
\begin{center}
\epsfig{file=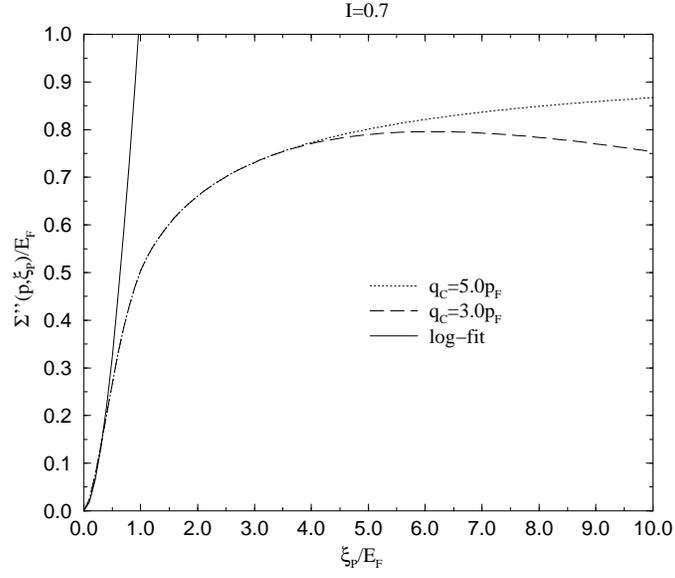,width=4in}
\end{center}
\caption{$\Sigma''(p,\xi_p)$ on a linear energy 
scale for $q_c=5.0p_f$ and $3.0p_f$ when ${\bar{I}=0.7}$.}
\label{seon}
\end{figure}

\begin{figure}
\begin{center}
\epsfig{file=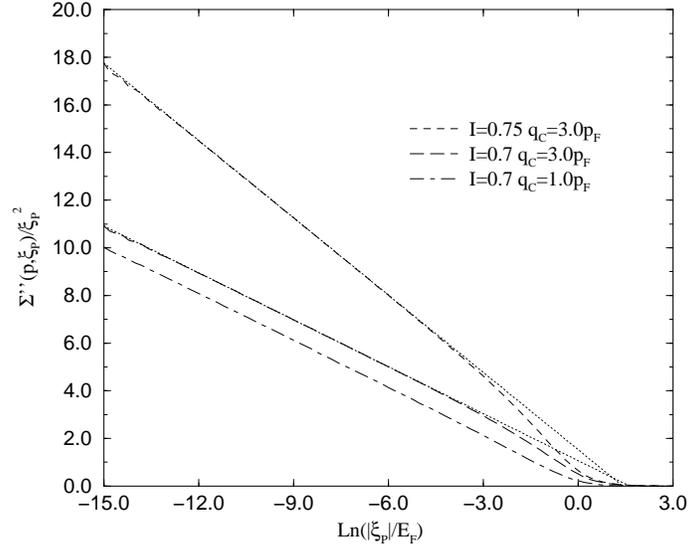,width=4in}
\end{center}
\caption{ $\Sigma''(p,\xi_p)$ on a logarithmic 
energy scale for parabolic band (for the same $q_c$'s but 
different $\bar {I}$'s the cut-offs, $\xi_0$, are the same. The 
energies are in the unit of $E_f$).}
\label{selnon}
\end{figure}

\begin{figure}
\begin{center}
\epsfig{file=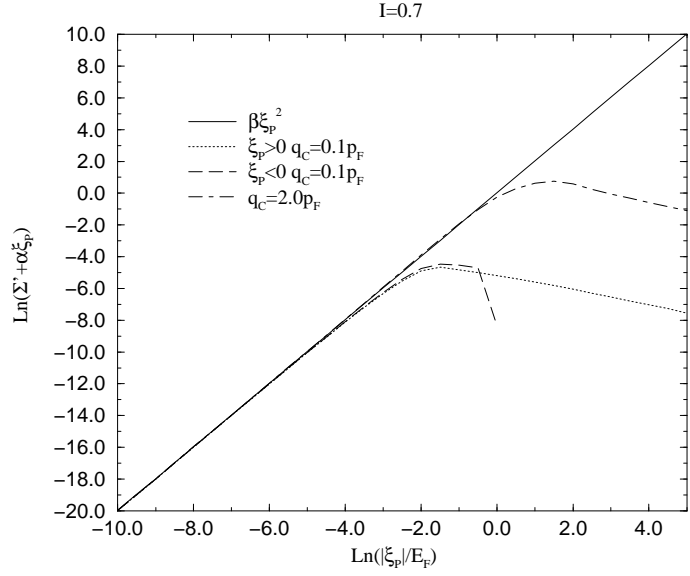,width=4in}
\end{center}
\caption{The ${\xi_p}|{\xi_p}|$ term in the real part of 
$\Sigma(p,\xi_p)$. (Energies are in the unit of $E_f$.)}
\label{ser2}
\end{figure}

 The leading dependence on $\xi_p$ in 
the imaginary part of self-energy in  
$\Sigma({\bf p},E)=\Sigma'({\bf p},E)+
\imath\Sigma''({\bf p},E)$ on shell is
\begin{equation}
\Sigma''({\bf p},\xi_{{\bf p}})=\beta
\xi^2_{\bf p} \ln |{\xi_{{\bf p}} \over \xi _{0} }|
+O(\xi_{\bf p}^4)
\end{equation}
where $\beta$ is independent of the direction for a parabolic band 
and given by
\begin{eqnarray}
\beta&=&-\int^1_0 {d\mu \over \sqrt{1-\mu^2}}
({1 \over 2}{A^3_s \over [(1-\mu^2)+(A_s\mu)^2]}+
{3 \over 2}{A^3_a \over [(1-\mu^2)+(A_a\mu)^2]})
\\ \nonumber \cr
&\approx&-{(1+{\bar{I}}+{\bar{I}}^2)\over {(1-{\bar{I}}^2)^2}} 
{{\bar{I}}^2 \over{4 E_f}},
\end{eqnarray}
$A_s={\bar I \over 1+\bar I}$, $A_a={\bar I \over 1-\bar I}$ and 
$\bar I=N(0)I$ where $N(0)=m/2\pi\hbar$ is the density of states at 
the Fermi surface for a 2D parabolic band for up or down spins.
 The real part of the self-energy for a parabolic band is
\begin{equation}
\Sigma'({\bf p},\xi_{\bf p})=-\alpha \xi_{{\bf p}}
-{\pi \over 2}\beta \xi_{{\bf p}}|\xi_{\bf p}|
-\gamma |{\xi}_{\bf p}|^{5/2}+O(\xi_p^3)
\end{equation}
where 
for the paramagnon model,
\begin{mathletters}
\begin{equation}
\alpha \approx {(2+\bar{I}) \over {(1-{\bar{I}}^2})}
{{\bar{I}}^2 min[q_{c},2p_f] \over {\pi p_f}}
\end{equation}
\begin{equation}
\gamma \approx {\bar{I} \over {4\pi E^{3/2}_f}}
\end{equation}
\end{mathletters}
where $E_f=p_f^2/2m$  and $\xi_0 \propto p_f$ for $q_c>2p_f$ 
and $\xi_0 \propto q_c$ otherwise \cite{Kim}. The logarithmic 
behavior is restricted to $|\xi_p|<\xi_0$.
 In the RPA approximation the largest contribution comes from 
the antisymmetric channel since the contribution of the channel is 
enhanced by a $1/(1-\bar I)^2$ factor in the case of a repulsive 
interaction. The generic FL behavior in $\Sigma''(p,\xi_p)$ is shown
 in the figure\ \ref{seon} with a log-fit. The figure\ \ref{selnon} 
shows an interaction $\bar{I}$ dependence in the logarithmic scale 
of $\xi_p$ and $q_c$ dependence. The cut-off energy $\xi_0$ depends 
only on $q_c$ and the slope $\beta$ depends on $\bar I$. The q-sum 
in the equation (4) has a limit of $min[q_c,2p_f]$, so that the 
log-fit for $q_c=3.0 p_f$ and $5.0 p_f$ has the same cut off in the 
log dependence. The log fit is good to within 10 percent up to 
$\xi_p=0.2E_f$ which, however, depends on $q_c$. The $\xi_p|\xi_p|$ 
term in $\Sigma'(p,\xi_p)$ is related to the 
$\xi_p^2 \ln |\xi_p|$ term in the imaginary part through the 
Kramers-Kronig relation\cite{Coffey} and also depends only on long 
wavelength properties. In the figure\ \ref{ser2} the 
${\xi_p}|{\xi_p}|$ term has been isolated in the logarithmic scale, 
so that the slope 2 indicates ${\xi_p}^2$ in 
$\ln|\Sigma'+\alpha {\xi_p}|$. This contribution comes from the 
dependence of $V^{eff}(q,\omega)$ on the variable 
$s=\omega/q v_f$\cite{footnote} 
where $s$ is the variable which 
appears in $\chi(\bf q,\omega)$ in the long-wavelength limit
\begin{equation}
\lim_{{\bf q} \rightarrow 0}\chi(q,\omega)
=\chi(s)=N(0)[-1+{s \over \sqrt{s^2-1}} \Theta(|s|-1)
+\imath {s \over \sqrt{1-s^2}} \Theta(1-|s|)].
\end{equation}
 As in the 
$\Sigma''(p,\xi_p)$, the coefficient is independent of $q_c$ and 
the log fit survives up to $\xi_p \approx \xi_0$. The term is present
 both for $\xi_p>0$ and $\xi_p<0$. The intercept point of vertical 
axis shows the coefficient ${\pi \over 2}\beta$ and the curve for 
$\xi_p <0$ ends at $\xi=-E_f$ in $\ln|\xi_p|$. 
 The $|\xi_p|^{5/2}$ term is the leading zero sound contribution. 
This term has a corresponding $(\xi-\xi_{th})^{3/2}$ 
term in $\Sigma''(p,\xi_p)$\cite{Kim}, where $\xi_{th}=(v_{zs}-v_f)$,
 determined by the velocity of the zero-sound mode, $v_{zs}=[
(1+\bar I)/\sqrt{1+2\bar I}]v_f$.

 In 3D the leading dependence in $\Sigma''(p,\xi_p)$ is $\xi_p^2$ 
and the leading correction to this , $|\xi_p|^3$, 
corresponds to the term 
$\xi_p^2 \ln \xi_p$ in 2D. The $|\xi_p|^3$ depends 
only on long wavelength behavior\cite{Baym} and has a corresponding 
$\xi_p^3 \ln |{\xi_p \over \xi_0}|$ term in $\Sigma'(p,\xi_p)$.
 The cut-off in the log, $\xi_0$, depends on both $\bar I=N(0)I$ and
 $q_c$ because of the stronger $q$ dependence in real part of 
$\chi^{3D}(q,\omega)$. The magnitude 
of these $\xi_p^2 \ln \xi_p$ and $\xi_p |\xi_p|$ terms in 2D and the 
$|\xi_p^3|$ and $\xi_p^3 \ln \xi_p$ terms in 3D are enhanced by 
repeated scattering of $ph$ pairs. The strength of the repeated 
scattering depends on the coupling constant, 
$\bar I$, which is independent of $p_f$ in 2D but goes to zero
 in 3D as $p_f \rightarrow 0$. As a result unlike the case in 3D, 
where the $ph$ channel is not important for low densities, the 
enhancement is independent of density in 2D and the results derived 
here for 2D remain the leading corrections even in low densities.
 However both in 2D and 3D $\Sigma(p,E) \rightarrow 0$ as the phase 
space for holes vanishes. The leading correction term of 2D FL is 
limited to a small energy region which vanishes as $p_f$ decreases 
in the low density limit.

\begin{figure}
\begin{center}
\epsfig{file=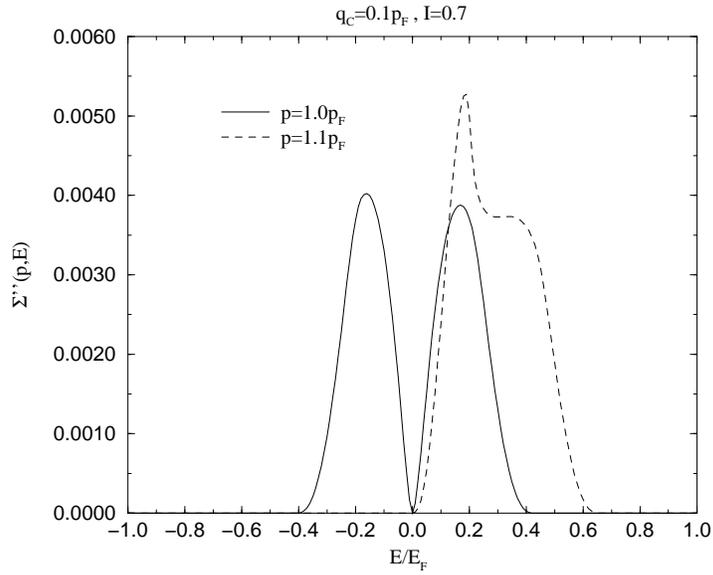,width=4in}
\end{center}
\caption{$\Sigma''(p,E)$, off-shell self-energy, 
for $p=p_f$ and $1.1p_f$. 
$\Sigma''(p,E)$ has a very different form from $\Sigma''(p,\xi_p)$.
 The structure of $\Sigma''(p,E)$ has $q_c$ dependence through step 
function in equation (4).}
\label{seioff}
\end{figure}

\begin{figure}
\begin{center}
\epsfig{file=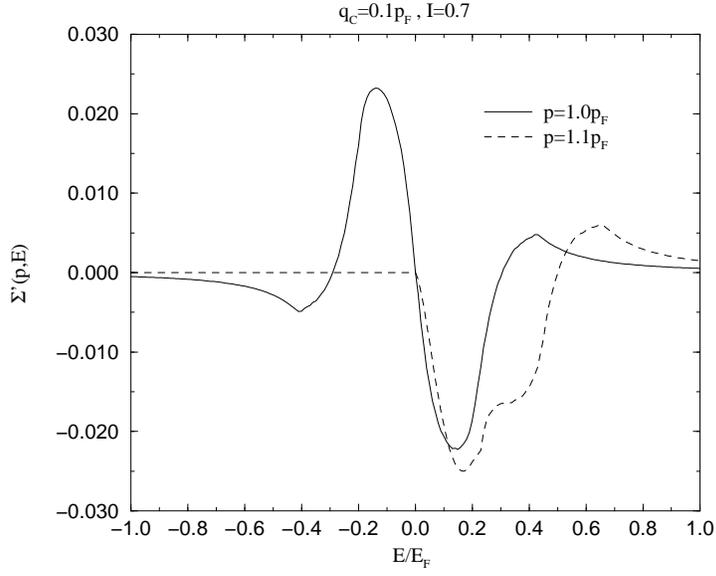,width=4in}
\end{center}
\caption{$\Sigma'(p,E)$, off-shell real part of self-energy, 
for $p=p_f$ and $1.1p_f$. The structure of $\Sigma'(p,E)$ also has 
$q_c$ dependence through step function in equation (4).}
\label{seroff}
\end{figure}

 The dependences of $\Sigma(p,E)$ on the variables, $p$ and $E$, 
are very different. For instance $\Sigma''(p,E)$ has a 
$E^2\ln {|E| \over E_\pm}$ behavior only for 
$p=p_f$, where $E_+=E_-=\xi_0$ are 
cutoffs for $E>0$ and $E<0$ respectively. In the 
figure\ \ref{seioff} and the figure\ \ref{seroff} we 
show $\Sigma'(p,E)$ and $\Sigma''(p,E)$ 
vs. E for $p={p_f}$ and ${1.1p_f}$. As $p$ goes away from $p_f$, the 
structure of the self-energy vanishes inside the Fermi surface 
(for $p<p_f$ the structure vanishes outside the surface) due to 
the step function in the zero temperature self-energy expression in 
the equation (4) and so there is a threshold in the self-energy for 
$|E|<|\xi_{{\bf p}-{\bf q}}|_{min}$. The sharp threshold effect is absent for 
interaction without a cutoff, $q_c$. However if $V^{eff}(q,\omega)$ 
falls off with increasing $q$ then there is an effective $q_c$ and 
associated with that there 
is an effective cutoff in $E$ beyond which $\Sigma(p,E)$ is reduced in 
magnitude. The most direct probe of the $E$ dependence
 of $\Sigma(p,E)$ is in angle resolved photoemission spectroscopy in 
which the spectral density is measured. The quasi-particle peak in 
the data depends on the $E$ dependence of $\Sigma(p,E)$ and has been 
analyzed for the cuprate superconductors in both the normal and 
superconducting state\cite{Shen1}. It has been found that the 
linewidth of quasi-particle peak does not vary as $E^2$ which has 
been pointed as not being the FL dependence\cite{Shen2}. 
The strong $p$ and $E$ dependence of $\Sigma(p,E)$ found here 
for a simple parabolic band shows that deviations from a simple 
$E^2$ dependence are to be expected. This is especially the case 
for cuprates with strong bandstructure effects and quasi-particle 
interactions giving rise to low energy spin 
fluctuations\cite{Kouznetsov}.

\subsection{Self-Energy at Finite Temperature}

\begin{figure}
\begin{center}
\epsfig{file=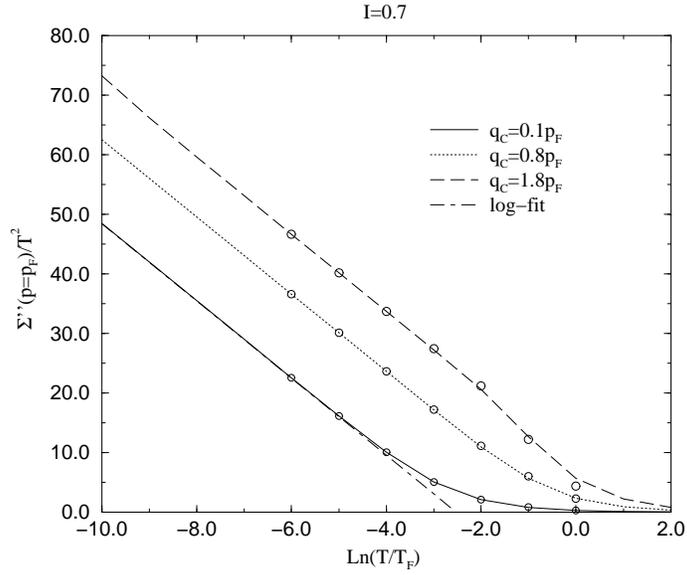,width=4in}
\end{center}
\caption{The temperature dependent $\Sigma''(p_f,\xi_{p_f})$ on a 
logarithmic T scale (The circles calculated by the temperature
 dependent $\chi''(q,\omega)$.)}
\label{tselnt}
\end{figure}

\begin{figure}
\begin{center}
\epsfig{file=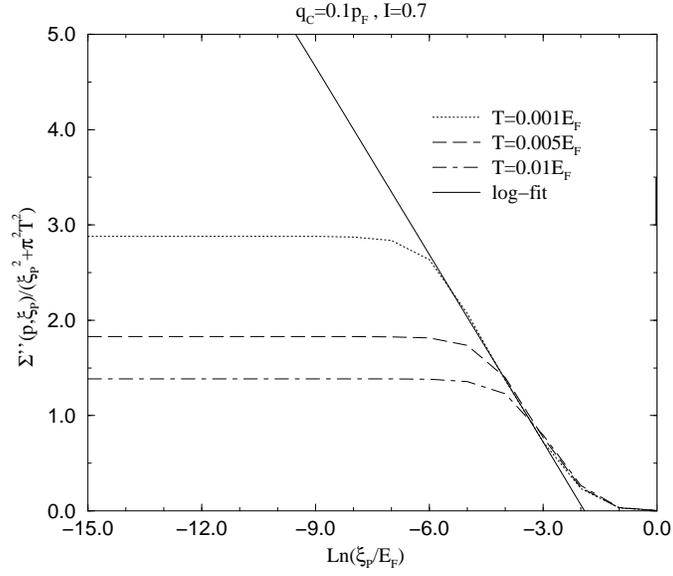,width=4in}
\end{center}
\caption{The temperature dependent $\Sigma''(p,\xi_p)$ on a 
logarithmic energy scale at $T=0.001E_f$, $0.005E_f$ and $0.01E_f$.}
\label{tseln}
\end{figure}

\begin{figure}
\begin{center}
\epsfig{file=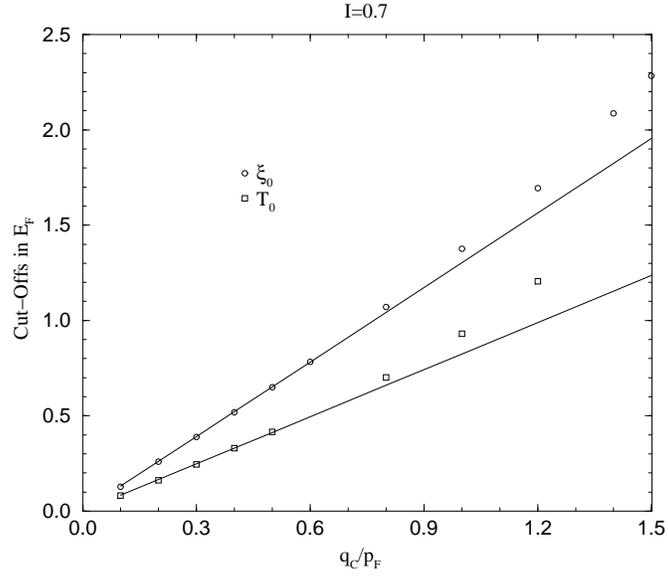,width=4in}
\end{center}
\caption{The cut-offs ; $\xi_0$ from $\Sigma''(p,\xi_p)$ at zero 
temperature and $T_0$ from $\Sigma(p=p_f,\xi_p,T)$, as $q_c$ 
approaches zero $\xi_0$ and $T_0$ have linear dependence in 
$q_c$.}
\label{cutoff}
\end{figure}

 At finite temperature $\Sigma''(p,\xi_p)$ is
\begin{equation}
\Sigma''({\bf p},\xi_{\bf p})=
-\sum_{\bf q}[f(-\xi_{{\bf p}-{\bf q}})
+n(\xi_{\bf p}-\xi_{{\bf p}-{\bf q}})]
Im V^{eff}({\bf q},\xi_{\bf p}-\xi_{{\bf p}-{\bf q}})
\end{equation}
where $n$ and $f$ are Bose and Fermi distribution functions 
respectively.
 The temperature dependence of 
$\Sigma''(p_f,\xi_{p_f})$ is plotted for 
different values of $q_c$ in the 
figure\ \ref{tselnt}. $\Sigma''(p,\xi_p)$ calculated with 
$\chi(q,\omega)$ evaluated at zero temperature is the solid lines 
and with the temperature dependent $\chi''(q,\omega,T)$ are the 
circles. As can be seen from figure\ \ref{tselnt}, the temperature 
dependence in the $\chi''(q,\omega,T)$ does not affect the cut-off 
$T_0$ since the $\chi(q,\omega,T)$ does not vary much respect to $T$
 for $T \ll E_f$ in the long wave length limit.  The leading 
dependence of $\Sigma''(p,\xi_p)$ on $p$ and $T$ is given by 
\begin{equation}
\Sigma''(p,\xi_p)=\beta(\xi_p^2+\pi^2 T^2) \ln(max[\xi_p,T]/T_0)
\end{equation}
which is shown on the figure\ \ref{tseln}. 
The zero temperature and finite temperature cutoffs, $\xi_0$ and 
$T_0$, are shown in figure\ \ref{cutoff}. They have similar values 
and increase for $q_c \leq 2p_f$ and are independent of $q_c$ for 
larger values of $q_c$. Although the self-energy at finite 
temperature does not have explicit step functions as in zero 
temperature case, the finite temperature cutoffs are independent of 
$q_c \geq 2p_f$ since the log behavior comes from low temperatures.
 Since both $\xi_0$ and $T_0$ are proportional to $p_f$ at low 
densities the leading FL behavior in the temperature case is also 
limited to a region which depends on the density of the system.

\begin{figure}
\begin{center}
\epsfig{file=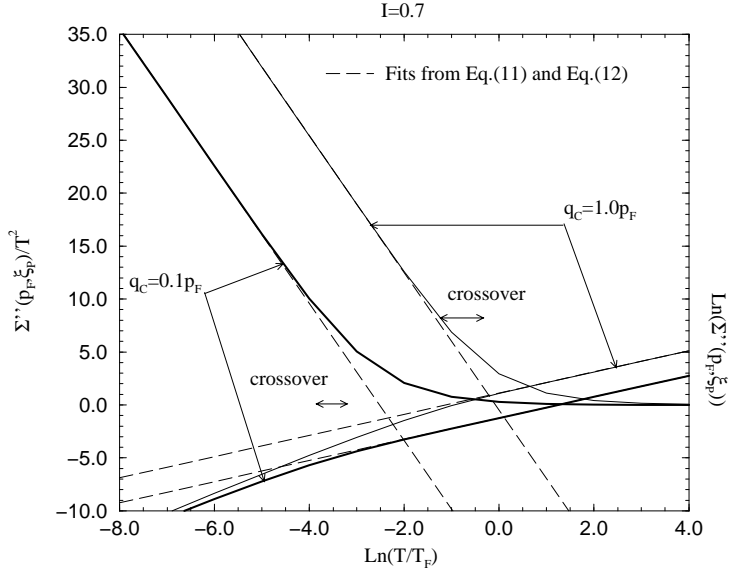,width=4in}
\end{center}
\caption{$T^2\ln T$ and linear $T$ dependence in 
$\Sigma''(p_f,\xi_p)$. The dotted lines are the fits from equations 
(11) and (12). The left solid line is $\Sigma(
p_f,\xi_p)/T^2$ and the right is $\ln[\Sigma(p_f,\xi_p)]$. The 
crossover from  the $T^2\ln T$ to $T$ behaviors has been marked 
by $\leftrightarrow$. The width of the cross region is 
$\sim0.01E_f$ for $q_c=0.1p_f$.}
\label{tstl}
\end{figure}

 As $T$ increases beyond the degenerate temperature region, the 
Bose distribution term dominates the contribution from the Fermi 
Dirac distribution term in the equation (10). In the high 
temperature limit $\Sigma''(p_f,\xi_{p_f})$ is approximated by
\begin{equation}
\Sigma''(p_f,\xi_{p_f}) \approx -\beta {8 min[q_c,2p_f] \over p_f}T 
+O({1 \over T})  
\end{equation}
 where the linear term in T comes from the Bose contribution. 
 $\Sigma''(p_f,\xi_{p_f})$ has been graphed in two ways, 
$\Sigma''(p_f,\xi_{p_f})/T^2$ vs. $\ln T$ and 
$\ln[\Sigma''(p_f,\xi_{p_f})]$ 
vs. $\ln T$ in the figure\ \ref{tstl} for $q_c=0.1p_f$ and 
$q_c=1.0p_f$.
 The dashed lines are the logarithmic and linear 
temperature dependence determined from equations (11) and (12).
 The size of the crossover region from the low 
temperature limit to the high temperature limit is small as can be 
seen in the figure and is from $T=0.02E_f/k_B$ to $T=0.03E_f/k_B$.
 This is estimated by considering 10 percent 
deviation from the fits. Although the linear $T$ dependence 
comes from the high temperature limit, as soon as the temperature 
dependence behavior leaves the $T^2\ln T$ behavior it tends to reach
 the linear temperature dependence rapidly. Since $T_0$ is 
proportional to density the 
leading FL behavior survives in a low temperature region which 
shrinks with decreasing density and the linear high temperature 
behavior is followed at lower and lower temperatures. The size of 
 crossover also decreases with density. The same behavior is seen in 
3D with similar crossover properties. Consequently this behavior is
 not determined by band-structure or low 
dimensionality and it may be misleading to refer to it as 
Luttinger-like\cite{Aristov}.

\subsection{Thermodynamic Properties}

\begin{figure}
\begin{center}
\epsfig{file=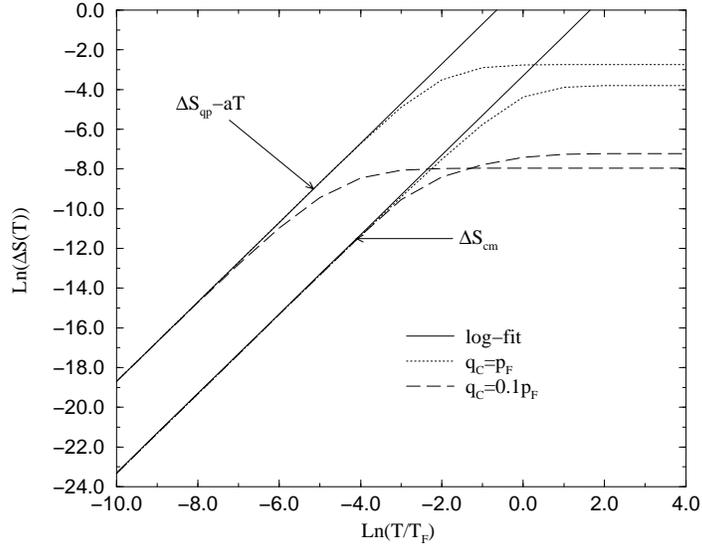,width=4in}
\end{center}
\caption{$T^2$ contributions in entropy from quasi-particle 
scattering with the continuum and with the collective mode in 
entropy; the solid line is the $T^2$ fit, the $T$ dependence term 
has been subtracted from $\Delta S_{qp}(T)$. The continuum 
contribution is $\sim 10^2$ larger than the zero sound contribution.}
\label{entln}
\end{figure}

 The contribution from quasiparticle interactions to the 
thermodynamic potential, $\Delta \Omega$, 
can be calculated by linked cluster expansion\cite{Mahan}. After 
analytic continuation to the real $\omega$ axis $\Delta \Omega$ can 
be split up into 
$\Delta \Omega_{qp}$ from a quasiparticle contribution and 
$\Delta \Omega_{cm}$ from a collective modes;
\begin{eqnarray}
\Delta \Omega(T)&=&-T\sum_{\imath \omega_n} \sum_{\bf q} \int^1_0 
{d \eta \over \eta} \chi(\eta,q,\imath \omega_n)V^{eff}
(\eta,q,\imath \omega_n)
\\ \nonumber \cr
&=&\Delta \Omega_{qp}+\Delta \Omega_{cm}.
\end{eqnarray}
First, $\Delta \Omega_{qp}$ is
\begin{equation}
\Delta \Omega_{qp}(T)=\sum_{|\bf q|<q_c} \int^\infty_0 
{d\omega \over \pi}
n(\omega)[F({\bf q},\omega)+2I^2\chi''({\bf q},\omega)],
\end{equation}
where
\begin{equation}
F(q,\omega)=\sum_{\lambda=s,a} \nu_\lambda 
tan^{-1}[{-V_\lambda \chi''(q,\omega) \over 1-V_\lambda \chi'
(q, \omega)}].
\end{equation}
 By taking a derivative $\Delta \Omega_{qp}$ respect to $T$, the 
shift in the entropy, $\Delta S_{qp}$, is
\begin{eqnarray}
\Delta S_{qp}(T)=&-&\sum_{|\bf q|<q_c}\int^\infty_0{\partial 
n(\omega)
\over \partial T} [F({\bf q},\omega)+2I^2\chi''(\bf q,\omega)]
\\ \nonumber \cr
&-&\sum_{|\bf q|<q_c}\int^\infty_0 {d\omega \over 2\pi}
n(\omega)[{\partial F({\bf q},\omega) \over \partial T}
+2I^2{\partial \chi''({\bf q},\omega) \over \partial T}]
\end{eqnarray}
 The second term is negligible since, as shown in the previous 
section, the 
temperature dependence in $\chi(\bf q,\omega)$ is weak in the 
long-wavelength limit. The temperature dependence in entropy mainly 
comes from the first term. The entropy from quasiparticle 
contribution is
\begin{equation}
\Delta S_{qp}(T)=\Gamma_1 T +\Gamma_2 T^2 +O(T^3),
\end{equation}
\begin{mathletters}
\begin{equation}
\Gamma_1={\pi \over 6T_f}(A_s+A_a) {q_c \over p_f}
\end{equation}
\begin{equation}
\Gamma_2=-{6 \zeta (3)n \over \pi T^2_F} \beta
\end{equation}
\end{mathletters}
where $n$ is density of particles and $\zeta$ is the Riemann zeta 
function\cite{Coffey}.
 The figure\ \ref{entln} shows $T^2$ dependence for $q_c=0.1p_f$ 
and $q_c=p_f$. The $q_c$ dependent linear $T$ term has 
been subtracted. As in the figure\ \ref{ser2}, the slope of the term 
does not depend on $q_c$. This term can also be calculated by 
including $\Sigma'(p,\xi_p)$ in the quasiparticle spectrum and using
 the expression for the entropy of non-interacting gas. The 
quasiparticle spectrum has the $\xi_p|\xi_p|$ term which is 
directly connected to $T^2$ term in the entropy\cite{Coffey}.
 This $T^2$ describes the corrections to the $\Gamma_1 T$ dependence
 in $\Delta S_{qp}$ to better than 10 percent up to a temperature 
$T_S$ for entropy in the figure\ \ref{entln} which follows the same 
trend as $T_0(\sim \xi_0)$ in the figure\ \ref{cutoff} but
 is smaller. 

 The collective mode contribution, $\Delta \Omega_{cm}(T)$, is given
 by the zero-sound pole in the 
symmetric channel in $V^{eff}({\bf q},\omega)$. Taking derivative 
respect to $T$, one has a $T^2$ dependence in $\Delta S_{cm}$. In 
the figure\ \ref{entln} $\Delta S_{cm}$ is compared with 
$\Delta S_{qp}$.  The contribution to $T^2$ from the collective mode
 is about a percent of the quasi-particle contribution.

 The calculations discussed so far using the paramagnon model have 
shown that FL behavior is limited to regions of low energy and 
temperature which shrink with decreasing density. The cutoffs, 
$\xi_0$ and $T_0$, are determined by the finite momentum dependence 
of the quasiparticle interaction. We now turn to the 2D electron gas
 which has recently been of experimental and theoretical interest.

\section{2D Electron Gas}

 Murphy \emph{et al}.\cite{Murphy} have compared their data on the 
tunneling conductance in 2D quantum well systems to the expression 
for the single particle life time expression derived by Giuliani and 
Quinn\cite{Giuliani} using long wavelength approximation and 
found quantitative agreement if they multiplied this expression by 
6.3. Jungwirth \emph{et al}.\cite{Jungwirth} have investigated the 
same calculation with an effective interaction suggested by 
MacDonald \emph{et al}.\cite{MacDonald} between fermions with a 
static long wavelength limit in $V^{eff}$ and they obtained good 
fits to the data. We calculate the imaginary part of self-energy for
 the quasi-particles in n-type doped GaAs using the RPA.

 Instead of applying static approximation calculated in RPA with 
local corrections in $V^{eff}(q,\omega)$, we
 use the frequency dependent effective interaction,
\begin{equation}
V_q^{eff}({\bf q},\omega)={I_q^2 \chi({\bf q},\omega) 
\over 1-I_q \chi({\bf q},\omega)}
\end{equation}
where $I_q=2\pi e^2/\epsilon_s q$ or $\bar I_q=2\pi e^2 N(0)
/\epsilon_s q=0.02$\AA$^{-1}/q$\cite{Murphy}($N(0)$ is density
 of states at Fermi level). For GaAs the static dielectric constant,
 $\epsilon_s$, is about 10 and $m^*/m=0.067$. The only free
 parameter is $p_f$. As pointed out in the discussion of the 
contact interaction the coefficient of the 
$(\xi_p^2+\pi^2 T^2)\ln(max[\xi_p,T]/T_0)$ dependence in 
$\Sigma''(p,\xi_p)$ is determined 
by the long wavelength properties of $V^{eff}(q,\omega)$. 

\begin{figure}
\begin{center}
\epsfig{file=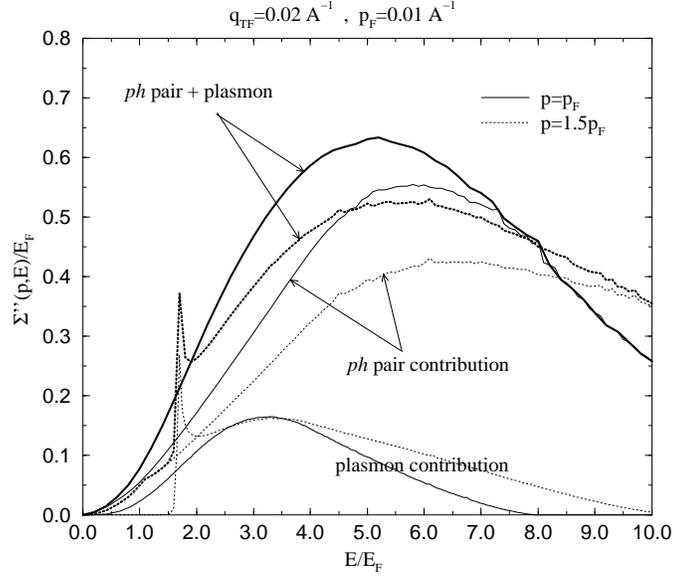,width=4in}
\end{center}
\caption{The incoherent \emph{ph} pair and plasmon contributions to 
$\Sigma''(p,E)$ for $p=p_f$ and $p=1.1p_f$.}
\label{plse}
\end{figure}

\begin{figure}
\begin{center}
\epsfig{file=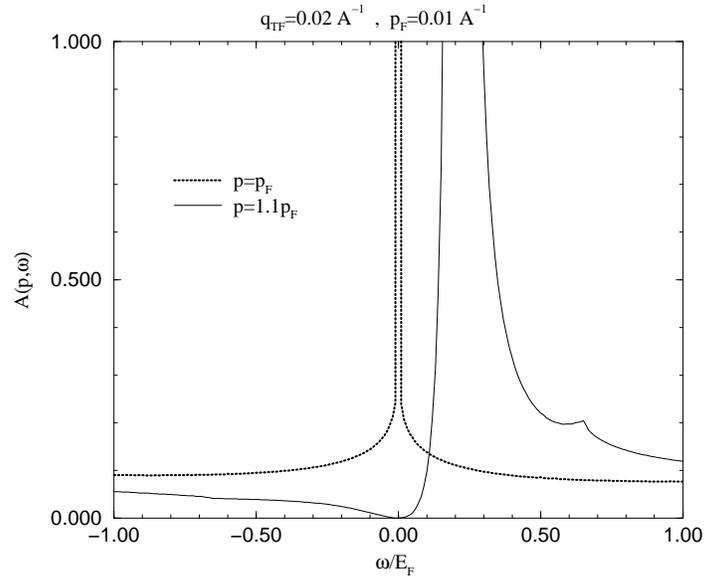,width=4in}
\end{center}
\caption{The spectral function, $A(p,\omega)$, when $p=p_f$ and 
$p=1.1p_f$. As $p$ is away from the Fermi surface, asymmetry occurs.
 $A(p,\omega)$ in the figure has \emph{ph} and plasmon contributions.
 The feature at $\omega/E_f \approx 0.6$ comes from the plasmon 
contribution.}
\label{specq}
\end{figure}

 There are two contributions to $\Sigma(p,E)$. One comes from the 
interaction of the quasi-particles with incoherent \emph{ph} pairs. 
This contribution gives the leading FL corrections and was discussed
 in detail above for the contact interaction. The second contribution
 comes from the interaction of quasi-particles with the plasmon mode.
 Although  it does not contribute to the leading $p$ 
and $E$ dependence of $\Sigma''(p,E)$, it is comparable in 
magnitude to that from the incoherent \emph{ph} pairs 
at finite $E$. These contributions and 
$\Sigma''(p,E)$ are plotted in figure\ \ref{plse} for $p=p_f$ and 
$p=1.1p_f$ at $T=0$. $\Sigma''(p_f,E)$ is symmetric and leads to a 
spectral density which is symmetric in the figure\ \ref{specq}. For 
$p \ne p_f$ $\Sigma''(p,E)$ is no longer symmetric in $E$ as was 
pointed out in the discussion of the contact interaction.

\begin{figure}
\begin{center}
\epsfig{file=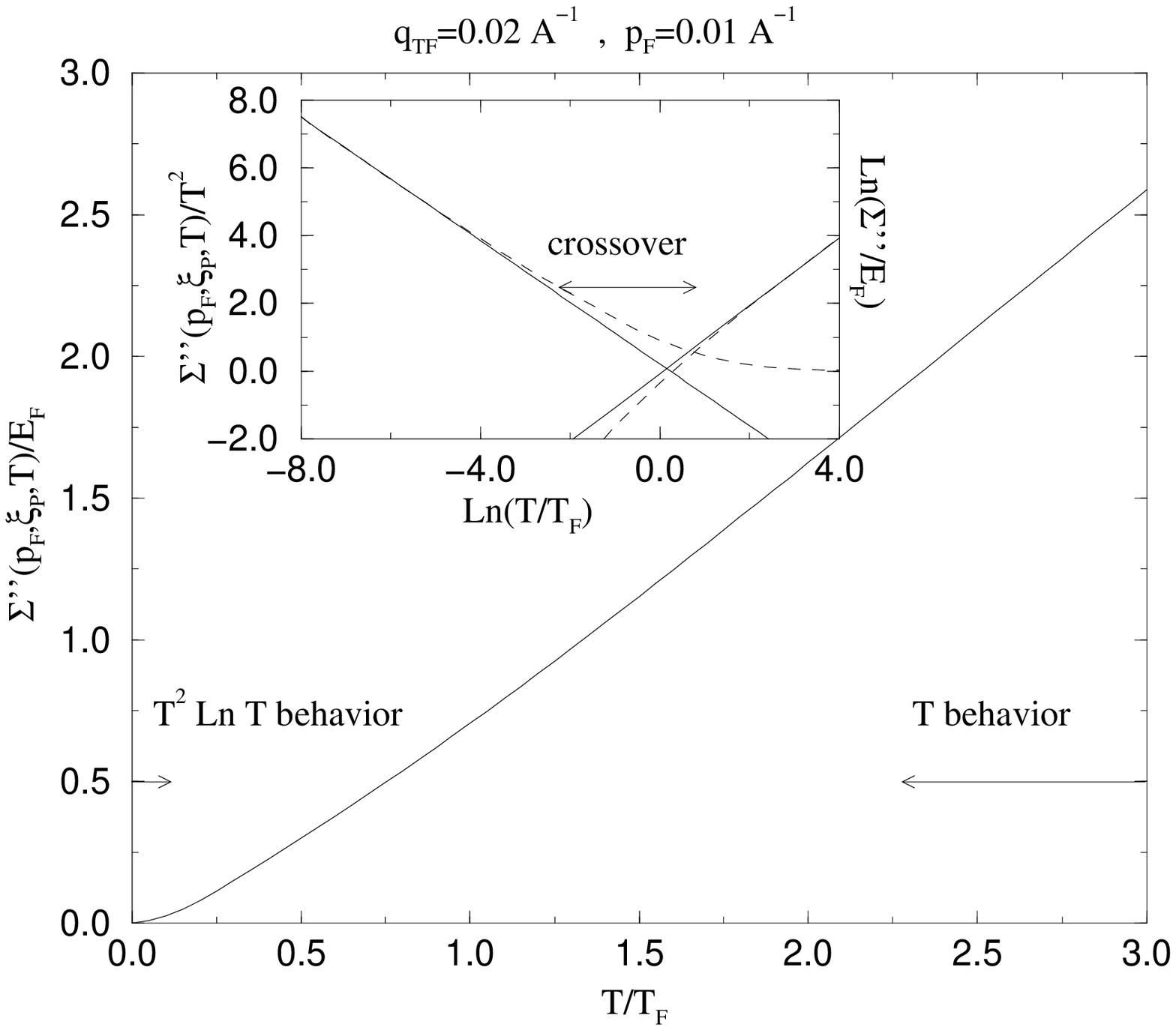,width=4in}
\end{center}
\caption{Extended plot of $\Sigma''(p_f,\xi_{p_f},T)$ for 2D 
quasi-particles for n-typed doped GaAs system with 
$n=1.6$x$10^{11}cm^{-2}$. Inset is plotted by logarithmic scale and 
shows crossover.} 
\label{qseit}
\end{figure}

 Using the parameters for quasi-particles in n-type doped GaAs 
quantum
 wells temperature dependence of $\Sigma''(p,E=\xi_p)$ is shown in 
figure\ \ref{qseit} for $n=1.6$x$10^{-11}cm^{-2}$ which gives 
$p_f=0.01$\AA$^{-1}$ and $E_f={p_f^2 \over 2m^*}=5.7 meV$.
 As in the contact interaction case there is a cross 
over from $T^2\ln T$ to linear $T$ behavior which depends on 
density. $T^2\ln T$ dependence is followed for $T_1<8 K$ and $T$ 
dependence for $T_2 > 156 K$ for $p_f=0.01$\AA$^{-1}$ system. 
These values of $T_1$ 
and $T_2$ were determined by requiring the $\Sigma''(p_f,\xi_{p_f})$
 to fit the $T^2\ln T$ and $T$ dependences to 10 percent. The 
$T^2\ln T$ dependence in the 
Coulomb interaction is also limited by low density limit as in 
contact interaction case. The cross over region ($T_1$ to $T_2$) has
 been estimated for different densities and is shown in table 1. The
 size of this region decreases with decreasing density system as in 
the contact interaction case but is larger.

 Quasiparticle interactions of an isotropic electron gas do not 
contribute to the transport relaxation time\cite{Mahan}. However the 
results of the calculations discussed here suggest that it is 
important to take into account the $p$ and $E$ dependences of 
$\Sigma''(p,E)$ and the plasmon contribution when comparing 
calculations of single particle properties in the 2D electron gas 
with experiment data.

\section{Conclusion and Summary}

We calculated the corrections to FL behavior in the single particle 
self-energies and in entropy using a contact and Coulomb 
interaction. We demonstrated that these corrections take the forms,
 $\xi_p|\xi_p|$ in $\Sigma'(p,\xi_p)$ and $\xi_p^2 \ln|\xi_p|$ in 
$\Sigma''(p,\xi_p)$, in a region close to the Fermi 
surface which shrinks with decreasing density and increasing 
temperature. In particular we have shown that at finite temperatures
 $\Sigma''(p_f,\xi_{p_f})$ rapidly goes over to 
a linear $T$ dependence 
even for a parabolic band. This type of linear 
temperature leads to a 
resistivity $\propto T$ which has been taken as evidence against a 
FL ground state in the cuprate superconductors. Our results suggest 
that an alternative explanation may be that the temperature below 
which FL behavior is seen is below the superconducting transition 
temperature. This has also been suggested by Aristov \emph{et al.
}\cite{Aristov} on the basis of band-structure effects in 2D Hubbard 
Hamiltonian.

 This point has been addressed by Boebinger 
\emph{et al}.\cite{Boebinger} who have applied magnetic fields to 
suppress superconductivity in transport measurements on the cuprates
 at low temperatures. It is claimed that this  reveals the 
normal state. They find that the non-FL behaviors persists. However 
recent work by Chan \emph{et al}.\cite{Chan} suggests that single 
particle properties of 2D electron systems are very sensitive to 
modest applied magnetic fields. They found that a pseudogap appears 
in the density of states in applied fields. A similar resistivity 
was also found in transport properties of 2D electron system in 
semiconductor hetero-structures by Simonian 
\emph{et al.}\cite{Simonian}. 
We will address the 
influence of magnetic fields on 2D FL in a forthcoming paper.

This work was supported by the New York State Institute for
Superconductvity(NYSIS).

%
%

%
%

\newpage
\begin{table}
\caption{Crossover temperatures from $T^2\ln T$ behavior ($T_1$)
to $T$ behavior ($T_2$) in the temperature dependence of 
$\Sigma''(p_f,\xi_{p_f})$ of the 2D electron gas. $T_1$ and $T_2$ 
are defined by 10 percent deviation from the fitting curves.} 
\begin{tabular}{ccc}
density (x$10^{11} cm^{-2}$)&$T_1$(K)&$T_2$(K)\\
\hline
6.37&25&231\\
3.58&20&236\\
3.12&16&258\\
1.93&12&185\\
1.59&8&156\\
\end{tabular}
\end{table}

\newpage

%
%

\begin{references}
\bibitem{footnote} In the second order calculation the 
$\xi_p|\xi_p|$ term exists for $p>p_f$ because it comes from 
$|s|=|p\mu/p_f| > 1$  in the real part of $\chi(s)$ where 
$\mu=\hat q \cdot \hat p$. However $|p\mu / p_f| < 1$  for $p<p_f$.
 As a result the on-shell self-energy calculated for second order 
can be a misleading indication of the $\xi_p$ dependence.
\end{references}
\end{document}